\documentclass[pdflatex,sn-nature]{sn-jnl}% Math and Physical Sciences Numbered Reference Style
\usepackage{graphicx}%
\usepackage{multirow}%
\usepackage{amsmath,amssymb,amsfonts}%
\usepackage{amsthm}%
\usepackage{mathrsfs}%
\usepackage[title]{appendix}%
\usepackage{xcolor}%
\usepackage{textcomp}%
\usepackage{manyfoot}%
\usepackage{booktabs}%
\usepackage{algorithm}%
\usepackage{algorithmicx}%
\usepackage{algpseudocode}%
\usepackage{listings}%
\theoremstyle{thmstyleone}%
\theoremstyle{thmstyletwo}%

\theoremstyle{thmstylethree}%

\begin{document}

\title[Article Title]{Quantum LiDAR with non-local modulation}

%%=============================================================%%
%% GivenName	-> \fnm{Joergen W.}
%% Particle	-> \spfx{van der} -> surname prefix
%% FamilyName	-> \sur{Ploeg}
%% Suffix	-> \sfx{IV}
%% \author*[1,2]{\fnm{Joergen W.} \spfx{van der} \sur{Ploeg} 
%%  \sfx{IV}}\email{iauthor@gmail.com}
%%=============================================================%%

\author*[1,2,3]{Xiao-Dong Fan}

\author[1,2]{Zhong-Hua Ou}\email{ozh@uestc.edu.cn}

\author[2,3,4]{Yun-Ru Fan}\email{yunrufan@uestc.edu.cn}

\author[5]{Kai Guo}\email{guokai07203@hotmail.com}

\author[6]{Zong-Liang Xie}

\author[1]{Qiang Qi}

\author[1]{Li-Xun Zhang}

\author[7]{Si Shen}

\author[3,7]{Hai-Zhi Song}

\author[3]{Yan-Yu Wei}

\author[8]{Hao Li}

\author[8]{Li-Xing You}

\author[2]{Qi Zhang}

\author[1]{Yong Liu}

\author[2,3,4,9]{Guang-Can Guo}

\author[1,2,3,4,9]{Qiang Zhou}\email{zhouqiang@uestc.edu.cn}

\affil*[1]{\orgdiv{School of Optoelectronic Science and Engineering}, \orgname{University of Electronic Science and Technology of China}, \orgaddress{\city{Chengdu}, \postcode{611731}, \country{China}}}

\affil[2]{\orgdiv{Center for Quantum Internet}, \orgname{Tianfu Jiangxi Laboratory}, \orgaddress{\city{Chengdu}, \postcode{641419}, \country{China}}}

\affil[3]{\orgdiv{Institute of Fundamental and Frontier Sciences}, \orgname{University of Electronic Science and Technology of China}, \orgaddress{\city{Chengdu}, \postcode{611731}, \country{China}}}

\affil[4]{\orgdiv{Key Laboratory of Quantum Physics and Photonic Quantum Information}, \orgname{University of Electronic Science and Technology of China}, \orgaddress{\city{Chengdu}, \postcode{611731}, \country{China}}}

\affil[5]{\orgdiv{Institute of Systems Engineering}, \orgname{AMS}, \orgaddress{\city{Beijing}, \postcode{100141}, \country{China}}}

\affil[6]{\orgdiv{Institute of Optics and Electronics}, \orgname{Chinese Academy of Sciences}, \orgaddress{\city{Chengdu}, \postcode{610209}, \country{China}}}

\affil[7]{\orgdiv{Southwest Institute of Technical Physics}, \orgname{}, \orgaddress{\city{Chengdu}, \postcode{610041}, \country{China}}}

\affil[8]{\orgdiv{National Key Laboratory of Materials for Integrated Circuits}, \orgname{Shanghai Institute of Microsystem and Information Technology}, \orgaddress{\city{Shanghai}, \postcode{200050}, \country{China}}}

\affil[9]{\orgdiv{CAS Center for Excellence in Quantum Information and Quantum Physics}, \orgname{University of Science and Technology of China}, \orgaddress{\city{Hefei}, \postcode{230026}, \country{China}}}

%%==================================%%
%% Sample for unstructured abstract %%
%%==================================%%

\abstract{Quantum light detection and ranging (LiDAR) utilizes quantum entanglement and correlation to improve precision, noise resilience and covertness of target detection. Despite recent advances, the development of a quantum LiDAR system that simultaneously achieves high precision and a large measurement range remains challenging. Here, we demonstrate a quantum amplitude-modulated continuous wave LiDAR with micrometer precision achievable via increased acquisition time and meter-scale measurement range. In our demonstration, the signal photons directly illuminate the target, while the idler photons are non-locally modulated with a high-frequency cosine wave and never interact with the target. By leveraging the non-local modulation and the quantum correlation, the target detection is achieved with a precision of 0.64 $\pm$ 0.06 mm within one second over a measurement range of 2-8 m. As the acquisition time is up to 500 s, the system achieves a precision of 29 $\pm\ 4{\ \mathrm{\mu m}}$. Furthermore, our system realizes a 50 times precision improvement over the classical single-photon scheme in a background noise 37 dB stronger than the returned probe photons. With these advantages, our method will open venues for the development of high-precision, long-range, and noise-resilient target detection.}

\maketitle

\section*{Introduction}
Light detection and ranging (LiDAR) has been widely used in various areas of science and technology \cite{RN3}, including autonomous driving \cite{RN51}, industrial process monitoring \cite{RN4}, and inter-satellite ranging \cite{RN14, RN2}. With the illumination of entangled/correlated photon pairs, quantum LiDAR achieves enhanced precision, noise resilience, and superior covertness, compared to conventional LiDAR in high-loss and high-background-noise environments \cite{RN6, RN11}. This novel method has attracted substantial interest in depth measurement \cite{RN15, RN17}, noise-resilience ranging and imaging \cite{RN70, RN12, RN13}, as well as covert ranging \cite{RN10}. 

The essential idea of quantum LiDAR is to exploit the quantum correlation between photon pairs. In such schemes, the signal photons are directed toward the target, whereas the idler photons do not interact with it, and then the information of the targets, such as distance and reflectivity, can be non-locally measured \cite{RN61, RN22}. Moreover, in the presence of background noise, the target detection sensitivity of quantum LiDAR can be improved by utilizing the quantum correlations in different degrees of freedom of photons, such as in intensity \cite{RN25, RN26}, spatial \cite{RN27, RN35, RN28}, spectral \cite{RN65, RN24}, polarization \cite{RN29, RN30}, and temporal domains \cite{RN8, RN33, RN31, RN139, RN66, RN69}. Despite these impressive results, the precision of the quantum LiDAR system, defined as the uncertainty in the target location \cite{RN37}, is limited to the centimeter or millimeter level due to the total time jitter of the quantum LiDAR system and the finite temporal sampling resolution. The quantum LiDAR with better precision can achieve more accurate target localization and enable a much broader range of applications. 
Using a single-photon detector (SPD) with ps-level timing jitter and a time-to-digital converter (TDC) with a sub-ps temporal sampling resolution increases the computational complexity of distance estimation and increases the cost of the system \cite{Boris2020-NP, McCarthy2025-optica}.
Significantly, a quantum LiDAR system that simultaneously achieves high precision and large measurement range has yet to be demonstrated, which is a key step toward practical quantum LiDAR.

The precision of quantum LiDAR can be improved to the micrometer level by using the quantum interference between the entangled/correlated photon pairs or undetected photons \cite{RN5, RN138, RN67, RN71}, such as the Hong-Ou-Mandel interferometer \cite{RN7}. This advantage has received increasing attention in the depth measurement of cell biology \cite{RN16, RN19}. 
However, the measurement speed of the quantum interferometer is limited by the slow mechanical motion device, and the measurement range is also limited by the optical path length of the reference arm. 
An alternative strategy to enhance the precision of quantum LiDAR is to employ the non-interferometric technique, such as electro-optic sampling \cite {Na-NP2020}, asynchronous optical sampling \cite {Dong-LPR2025, Jiang2024-PR}, and amplitude-modulated continuous-wave (AMCW) methods \cite{RN43, RN44, RN141}.
Among these, the AMCW method is more promising for developing a high-precision quantum LiDAR without the complex optical configuration.
In this protocol, AMCW LiDAR illuminates the target with amplitude-modulated light and determines the distance by measuring the phase difference between the transmitted and received modulated signals \cite{Li2024-PR}. The modulation waveform can be rapidly required through electronic sampling. By modulating a high-frequency cosine signal, AMCW LiDAR achieves micrometer-level precision \cite{RN43, RN141}. 
However, distances differing by integer multiples of the modulation period cannot be unambiguously resolved when the phase difference is larger than $2\pi$, thus restricting the measurement range to a single modulation period. Moreover, the use of coherent illumination fundamentally limits the precision in noisy environments and the covertness of conventional AMCW LiDAR systems.

In this work, we propose a quantum AMCW LiDAR that achieves target detection with simultaneous micrometer-level precision, meter-scale measurement range, and enhanced noise resilience, despite system timing jitter on the order of hundreds of picoseconds.
With this approach, the signal photons directly illuminate the target, and the idler photons are non-locally modulated with a single high-frequency cosine wave. 
The correlation between the signal and the idler photons is analyzed, as well as the relative distance between the quantum light source and the target can be estimated from the cosine-shaped correlation curves and the coincidence histograms. 
Unlike conventional AMCW LiDAR, the modulated idler photons never interact with the target, while we obtain target distance with high precision and extended measurement range in noisy environments through the non-local modulation and the quantum correlation. Quantum AMCW LiDAR simultaneously overcomes phase ambiguity, and the measurement range is not constrained by the modulation period.

In our demonstration, we reveal the physical mechanism of non-local modulation using the quantum theory of light, showing that non-local modulation curves such as cosine waveforms are reconstructed via temporal quantum correlations and their phases are proportional to the target distance. Theoretical precision is derived based on the Poisson distribution of photon counting and the Cramér–Rao bound.
Then, we experimentally demonstrate the quantum AMCW LiDAR by using a fiber-integrated quantum light source at the telecom band and a fiber-coupled confocal telescope system. When the modulation frequency is set to 1 GHz, we achieve a precision of 0.64 $\pm$ 0.06 mm with an acquisition time of one second within the measurement range from 2 to 8 m. As the acquisition time is up to 500 s, the system achieves a precision of 29 $\pm$ 4 ${\mathrm{\mu m}}$. When the signal-to-background noise ratio (SBR) is lower than -37 dB, the system with our method still exhibits a precision of 1.2 $\pm$ 0.1 mm within one second, which is over 50 times better than that of single-photon AMCW LiDAR without quantum correlation. This work will further broaden the development and applications of quantum LiDAR.

\begin{figure*}[h!]
\centering
\includegraphics[width=12 cm]{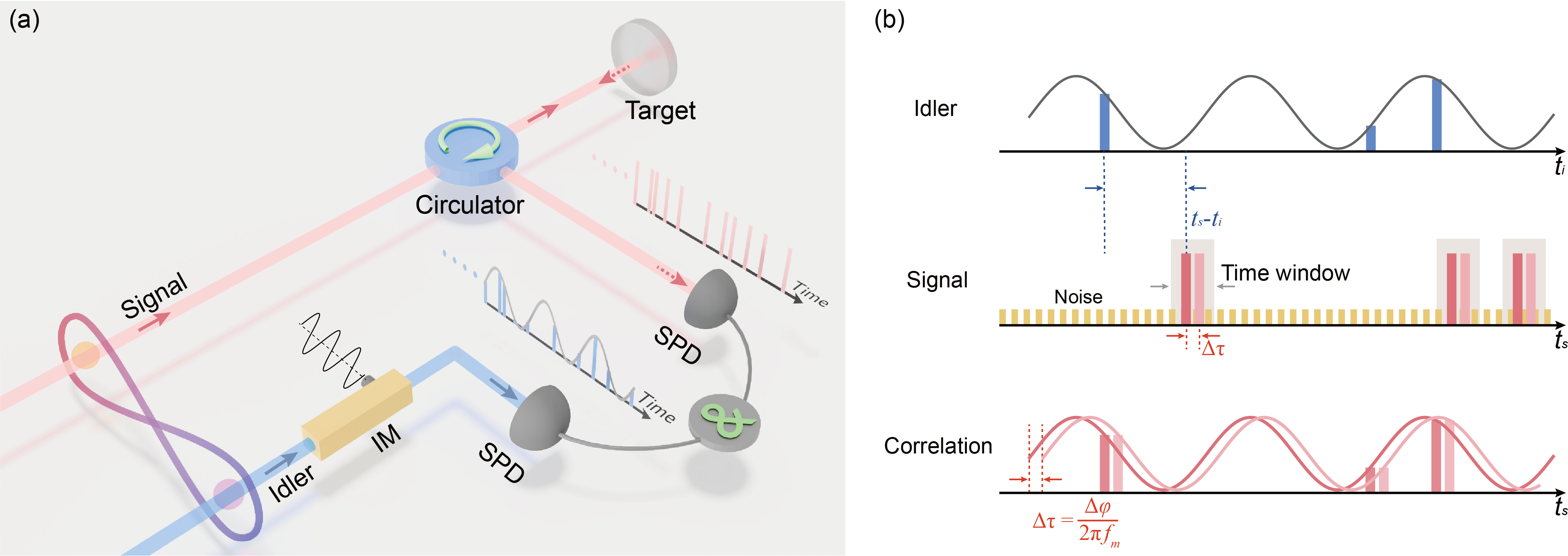}
\caption{Basic concept of the quantum AMCW LiDAR. (a) In the measured arm, the random signal photons propagate over a distance $d$ to the target, then are reflected by the target and detected by the SPD. In the herald arm, the idler photons are modulated into a cosine shape by the intensity modulator (IM) and detected by the SPD. The output electrical signals of SPDs are sent to a TDC for performing the temporal correlation analysis. (b) The timing diagram of the temporal correlation in three operational periods. During the period of the modulation signal, a signal photon event is selected in a time window with a fixed delay time when an idler photon event is heralded. Through the temporal correlation analysis, the reconstructed correlation curve matches that of the idler photons and shifts with the delay of signal photons. The coarse relative time $t_s - t_i$ can be measured by the coincidence histogram between the signal and idler photons. The fine relative time $\Delta \tau$ is extracted from the phase shift $\Delta \varphi$ of the reconstructed correlation curves. }
\label{fig:Fig1}
\end{figure*}

\section*{Results}
\subsection*{Principle}
Figure \ref{fig:Fig1}(a) shows the basic concept of quantum AMCW LiDAR. Correlated photon pairs can be generated via the nonlinear process. Their quantum state can be expressed as \cite{RN55, RN54}
\begin{equation}
    |\Psi \rangle =\int_{-\infty }^{\infty }{d{{t}_{s}}d}{{t}_{i}}\delta ({{t}_{s}}-{{t}_{i}})\hat{a}_{s}^{\dagger }\left( {{t}_{s}} \right)\hat{a}_{i}^{\dagger }\left( {{t}_{i}} \right)|vac\rangle 
\end{equation}
where the indices $s$ and $i$ indicate the signal photons (with time  $t_s$) and idler photons (with time $t_i$),  $\hat{a}_{s}^{\dagger}$ and $\hat{a}_{i}^{\dagger}$ are the creation operators of the signal and idler photons at time $t_s$ and $t_i$. The signal and idler photons are distributed to the measurement arm and the heralding arm, respectively. In the measurement arm, a target is placed at a distance of $d$. The signal photons are transmitted to the target and reflected, then detected by an SPD with a relative time delay $\tau$ (equals to $2d/c$, $c$ is the speed of light in vacuum). The positive-frequency part of the electric field operator for the signal photons can be expressed as
\begin{equation}
    E_{s}^{\left( + \right)}\left( {{t}_{s}} \right)=\int_{-\infty }^{\infty }{d{{\omega }_{s}}\sqrt{{{\alpha }_{s}}}{{{\hat{a}}}_{s}}\left( {{\omega }_{s}} \right){{e}^{-i{{\omega }_{s}}\left( {{t}_{s}}+\tau  \right)}}}
\end{equation}
where, $\omega_s$ is the angular frequency of the signal photons, $\alpha_s$ is the transmission coefficient of the signal photons in the measurement arm, ${{\hat{a}}_{s}}\left( {{\omega }_{s}} \right)$ is the annihilation operator for the signal photons at the angular frequency $\omega_s$. In the heralding arm, the temporal profile of the idler photons is cosine modulated by an intensity modulator and subsequently detected by another SPD. The positive-frequency component of the idler field operator at time ${t_i}$  can then be expressed as
\begin{equation}
    E_{i}^{\left( + \right)}\left( {{t}_{i}} \right)=\int_{-\infty }^{\infty }{d{{\omega }_{i}}\sqrt{\frac{{{\alpha }_{i}}}{2}\left[ 1+V\cos \,\left( 2\pi {{f}_{m}}{{t}_{i}}+{{\varphi }_{0}} \right) \right]}{{{\hat{a}}}_{i}}\left( {{\omega }_{i}} \right){{e}^{-i{{\omega }_{i}}{{t}_{i}}}}}
\end{equation} 
where, $\omega_i$ is the angular frequency of the idler photons, $\alpha_i$ is the transmission coefficient of the idler photons in the heralding arm, ${{\hat{a}}_{i}}\left( {{\omega }_{i}} \right)$ is the annihilation operator of the idler photons at the angular frequency $\omega_i$, $f_m$ is the modulation frequency, $V$ is the visibility of the modulation signal, and $\varphi_0$ is the initial phase. 

Then, the temporal probability distribution for jointly observing the returning signal photons and the modulated idler photons can be calculated by using the second-order correlation function \cite{RN54}. And the second-order correlation function $G^{(2)}$ can be expressed as (see derivations in the supplementary material):
\begin{equation}
    {{G}^{\left( 2 \right)}}\left( {{t}_{s}},{{t}_{i}},\tau  \right)=\frac{1}{2}{{\alpha }_{s}}{{\alpha }_{i}}\left[ 1+V\cos \,\left( 2{{f}_{m}}{{t}_{i}}+{{\varphi }_{0}} \right) \right]\delta \left( {{t}_{i}}-{{t}_{s}}-\tau  \right)
\end{equation}      
The $G^{(2)}$ can be further rewritten as a function of the signal photon arrival time $t_s$ :
\begin{equation}
    {{G}^{\left( 2 \right)}}\left( {{t}_{s}},\tau  \right)\propto 1+V\cos \left( 2\pi {{f}_{m}}{{t}_{s}}+\Delta \varphi +{{\varphi }_{0}} \right)
    \label{eq5}
\end{equation}        
where $\Delta \varphi = 4\pi {{f}_{m}}d/c$ is the phase shift induced by the target displacement. According to Eq. (\ref{eq5}), the temporal correlation curve has the profile of the modulation curve in the idler channel and is delayed by time of the signal photons reflected from the target in a nonlocal way. As shown in Fig. \ref{fig:Fig1}(b), moving the target induces a phase shift that appears as a lateral displacement of the temporal correlation curve. Consequently, the round-trip delay of the signal photons can be extracted from the temporal correlation curve, despite the target never interacting with the modulated idler photons. 

Furthermore, assuming a narrow-band modulation signal, the precision of quantum AMCW LiDAR is determined by the modulation frequency and the uncertainty of the estimated phase: 
\begin{equation}
    \sigma (d)=\frac{c}{4\pi {{f}_{m}}}\sigma (\varphi )
\end{equation}
The maximum achievable modulation frequency will be primarily limited by the total time jitter of the system, which includes the photon lifetime, the time-bin width of the TDC, and the time jitter of the SPD. The phase is estimated by using the maximum likelihood estimation method in our scheme. At best, the uncertainty of the estimated phase can be characterized by the Cramér–Rao bound \cite{RN38}. Accounting for the Poisson distribution of photon counting, the dominant contribution to the uncertainty arises from the finite number of correlated photon pairs. Thus, the optimal precision is further written as (see details in the supplementary material):
\begin{equation}
    \sigma (d)=\frac{c\sqrt{2}}{4\pi V{{f}_{m}}\sqrt{{{N}_{cc}}}}
\end{equation}
where $N_{cc}$ is the number of correlated photon pairs. In addition, quantum AMCW LiDAR leverages quantum correlations to extend the unambiguous measurement range. In conventional AMCW LiDAR \cite{RN43, RN44}, the measurement range is fundamentally limited by the modulation frequency, since distances differing by integer multiples of the modulation wavelength cannot be distinguished. In our quantum scheme, the coarse relative distance can be measured by the position of the coincidence histogram between signal and idler photons \cite{RN33}. The maximal measurement range is determined by the maximum resolvable relative delay of the TDC, which is independent of the modulation period.

Quantum AMCW LiDAR is also inherently robust against environmental background noise. By exploiting quantum correlations between signal and idler photons, the correlation measurement enhances the distance accuracy even under strong background noise. In our post-processing, the signal photon event is selected in a time window when the idler photon event is heralded. Temporal correlation is exploited to precisely select signal photon events while filtering out uncorrelated noise photons outside the time window, as shown in Fig. \ref{fig:Fig1}(b). The time window width is typically set to an integer multiple of the modulation period to suppress the phase interference from the uncorrelated background noise photons, see details in the supplementary material. Furthermore, the theoretical precision of quantum AMCW LiDAR under noise can be given by : 
\begin{equation}
    \sigma (d)=\frac{c\sqrt{2}}{4\pi V{{f}_{m}}\sqrt{{{N}_{cc}}}}\sqrt{1+\frac{{{N}_{i}}\Delta {{\tau }_{w}}}{SBR}}
    \label{eq8}
\end{equation}
where the $N_i$ is the photon counting rate of idler photons, the $\Delta {{\tau }_{w}}$ is the time window width, and the SBR is the ratio between the number of correlated photon pairs and that of noise photons. Although the noise photons may be accidentally detected and falsely identified as correlated with idler photons, they are effectively suppressed by the time window. For example, when ${{N}_{i}}=1\text{ }{\ \mathrm{MHz}}$, $\Delta {{\tau }_{w}}=1\text{ }\mathrm{ns}$, the background noise in the quantum AMCW LiDAR can be reduced by three orders of magnitude. For the single-photon AMCW LiDAR without quantum correlation, all noise photons are detected without suppression, and the corresponding precision is the special case of Eq. (\ref{eq8}) as ${{N}_{i}}\Delta {{\tau }_{w}}=1$. Consequently, the precision of quantum AMCW LiDAR is superior to that of single-photon AMCW LiDAR in the presence of noise.

\begin{figure*}[h!]
\centering
\includegraphics[width=12 cm]{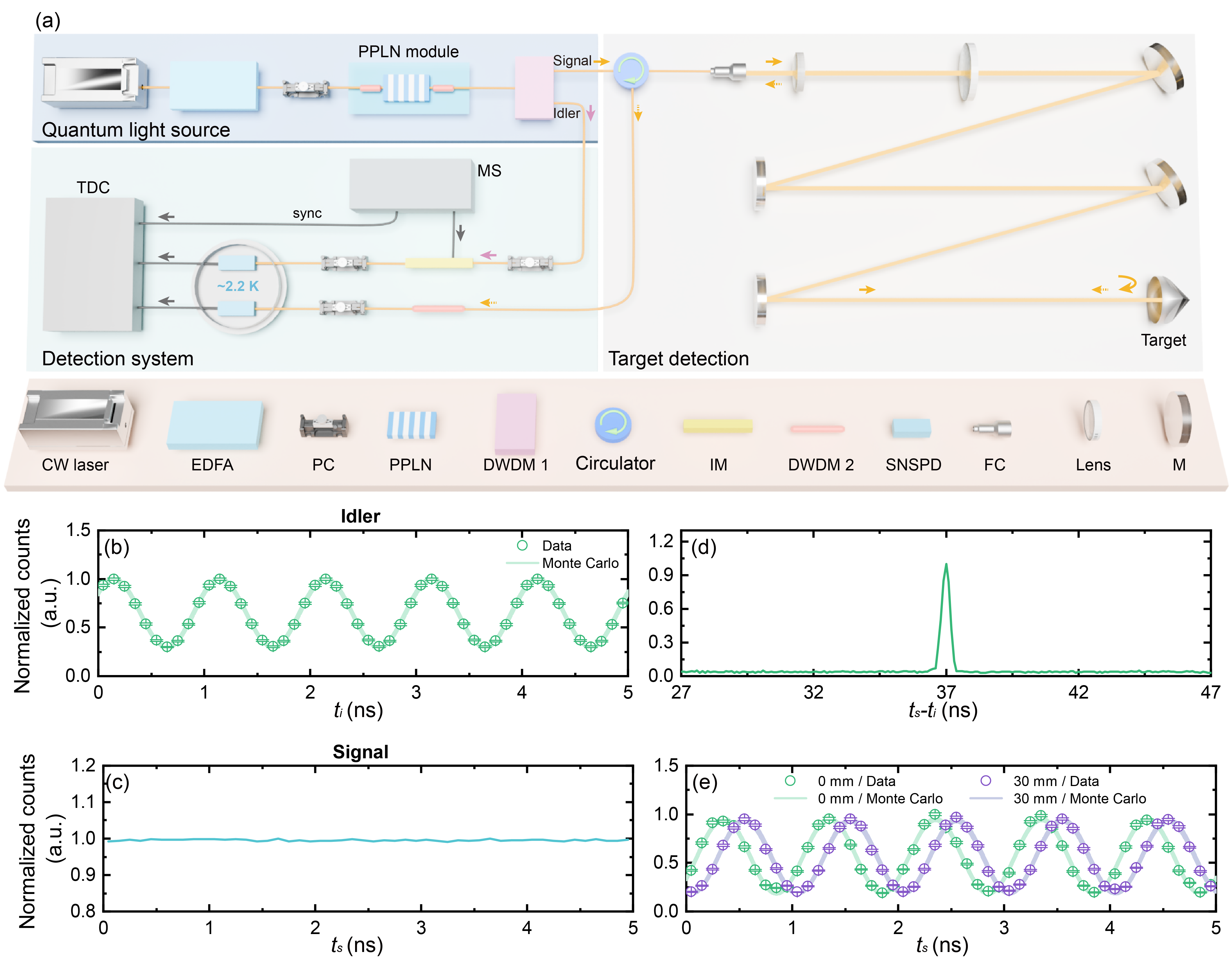}
\caption{Experimental setup and characterization of the quantum AMCW LiDAR. (a) Experimental setup. CW, continuous wave; EDFA, Erbium-doped fiber amplifier; PPLN, periodically poled Lithium Niobate; PC, polarization controller; DWDM, dense wavelength division multiplexer; FC, fiber collimator; M, mirror; IM, intensity modulator; MS, microwave source; SNSPD, superconducting nanowire single-photon detector; TDC, time-to-digital converter. (b) Waveform of the idler photons with the modulation frequency of 1 GHz. The counts are accumulated over a 5-ns period, and are normalized by the maximal counts. The green lines are the superimposed 1000-time Monte Carlo simulations. (c) Waveform of the signal photons returned from the target. (d) The coincidence histogram between the idler and signal photons. (e) The reconstructed correlation curves when the target is moved. The phase shift is caused by the relative displacement of the target, and the time window width is set as 1 ns. The green and purple lines are the superimposed 1000-time Monte Carlo simulations.}
\label{fig:Fig2}
\end{figure*}

\subsection*{Experimental setup}
Figure \ref{fig:Fig2}(a) illustrates the experimental setup of the quantum AMCW LiDAR system, which comprises a high-performance fiber-integrated quantum light source, target detection module, AMCW signal modulation electronics, and single photon detection module. The quantum light source generates the correlated photon pairs at 1.5 $\mathrm{\mu m}$ via cascaded second harmonic generation (SHG) and the spontaneous parametric down-conversion (SPDC) processes in a single piece of fiber-pigtailed PPLN waveguide \cite{RN45}. Continuous-wave pumping of this PPLN device, combined with manual tuning of the pump laser’s frequency and polarization, enables efficient generation of photon pairs with high spectral purity. By fiber-integrating the PPLN waveguide with noise-rejecting filters, spontaneous Raman scattering (SpRS) noise photons are significantly suppressed. Signal and idler photons at 1531.90 nm and 1549.32 nm, respectively, are selected by using dense wavelength-division multiplexers (DWDMs).

The signal and idler photons are then routed to the target detection module and the AMCW signal modulation section, respectively. The signal photons are coupled into free space via a fiber collimator. A fiber-coupled confocal telescopic system, comprising two lenses with different apertures and focal lengths, is used for both transmission and reception of the signal photons. These photons illuminate a target in free space, and the reflected photons are collected by the same telescopic system. The returning signal photons are detected by a superconducting nanowire single-photon detector (SNSPD), with an optical circulator and a polarization controller employed to optimize coupling and polarization alignment. Meanwhile, the idler photons are sent to an intensity modulator (with the half-wave voltage $V_{\pi}$ of 3 V, the bandwidth of 40 GHz), and their temporal intensity profile is shaped into a cosine waveform with a modulation frequency of 1 GHz. The modulated idler photons are directly detected by another SNSPD. The visibility is maximized by adjusting the voltage of the modulation signal, the bias voltage of the intensity modulator, and the polarization controller placed before the intensity modulator. The arrival times of both signal and idler photons are recorded by a TDC, enabling temporal correlation analysis. Figure \ref{fig:Fig2}(b) and (c) show the temporal distribution of signal and idler photons accumulated over 5 ns, respectively.

Then, the correlation curves can be measured via the temporal correlation analysis. Figure \ref{fig:Fig2}(d) shows the coincidence histogram between the signal and idler photons. Due to the correlation between photon pairs, there is a constant time difference between the arriving time of signal and idler photons. The delay time of 37 ns is caused by the length of the transmission fiber. The returned signal photon events are selected by the arriving time of idler photons with a fixed delay time and a 1-ns time window. The correlation curve of correlation photon-pairs over one modulation period is reconstructed, with the system synchronized to a 10 MHz reference clock that is phase-locked to the modulation signal, as illustrated in Fig. \ref{fig:Fig2}(e). The green and purple circles are the number of the correlation photon pairs after the temporal correlation analysis, and the green and purple curves are the fitting curves using a 1000-time Monte Carlo method. The reconstructed correlation curve retains the modulation frequency of the idler photons, with a phase shift caused by the round-trip time delay of the signal photons reflected from the target.

\subsection*{Distance measurement}
\begin{figure*}[h!]
\centering
\includegraphics[width=12 cm]{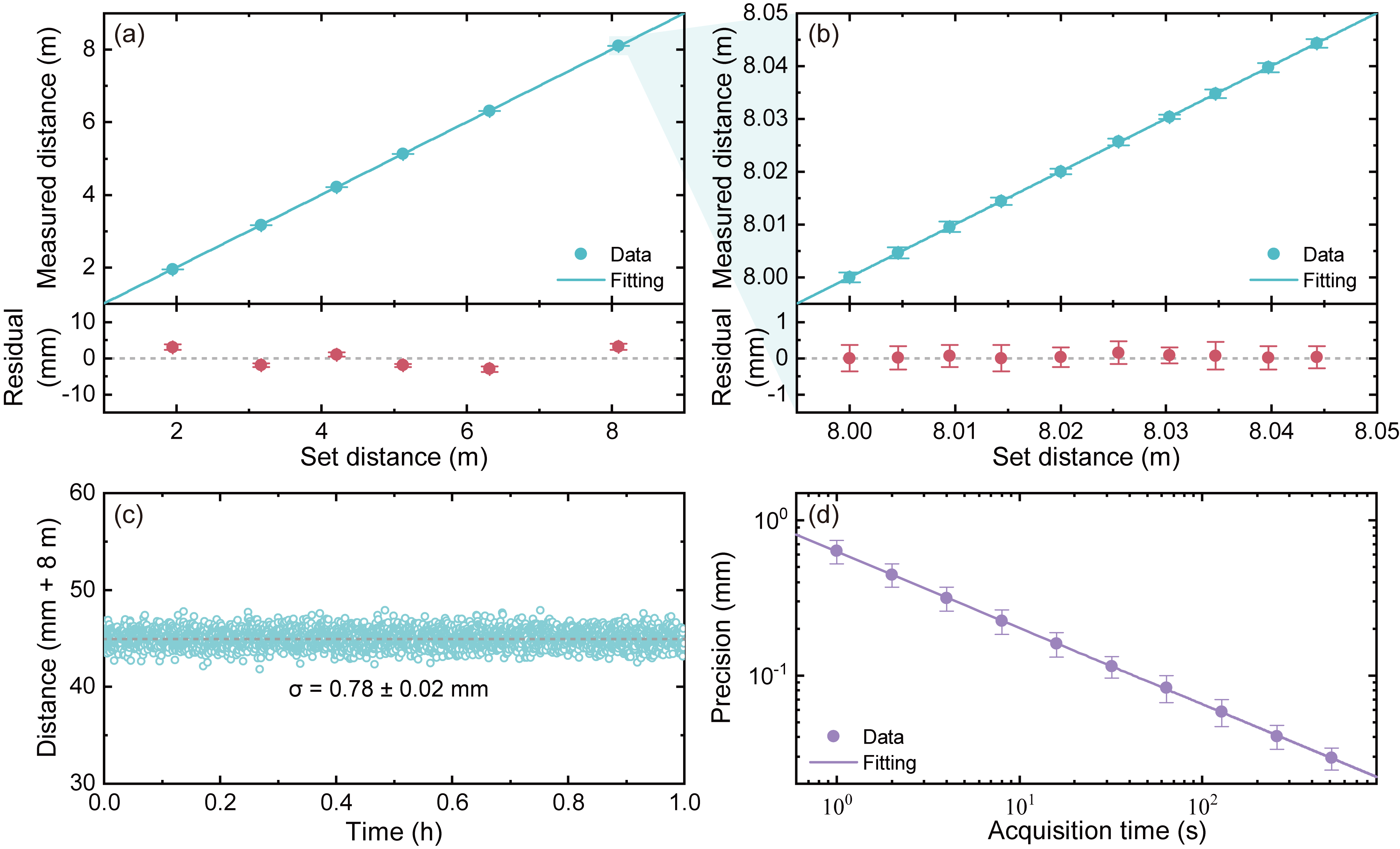}
\caption{Precision and measurement range of quantum AMCW LiDAR. (a) The distance measurement from 2 m to 8 m. (b) The distance measurement from 8 m to 8.045 m with 5-mm steps. Top, the distances estimated from the reconstructed correlation curves. Bottom, the residual and standard deviation at each measured distance. (c) Results of multiple measurements within 1 hour with the acquisition time of 1 s and the precision of 0.78 $\pm$ 0.02 mm. (d) Precision of quantum AMCW LiDAR at the distance of 8 m versus the acquisition time when the number of correlated photon-pair events is 14 kHz and the SBR is -23 dB. The precision follows the theoretical precision of Eq. (\ref{eq8}).}
\label{fig:Fig3}
\end{figure*}

Based on the experiment system described above, the ranging measurements are performed at both short and long distances. In the long-range regime, targets located beyond the distance corresponding to one modulation cycle (spatial distance of 15 cm corresponding to 1-GHz modulation frequency) can be coarsely localized by identifying the coincidence peak between signal and idler photons. When the target is positioned from 2 m to 8 m, which is over 50 times the unambiguous range, the absolute distance is successfully retrieved and exhibits a linear response over the entire interval, as shown in Fig. \ref{fig:Fig3}(a). The retrieved distance demonstrates excellent linearity with respect to the true distance, yielding a slope of 1.05 $\pm$ 0.02 and a coefficient of determination of $\mathrm{R^2}$ = 0.998. The absolute error is caused by the reference ruler. In the short-range measurement, the target is positioned at 8 m and scanned over a range of 0-45 mm (from 8 m to 8.045 m) with 5-mm steps. The measured distances still show a high linear relationship with the actual displacement, as depicted in Fig. \ref{fig:Fig3}(b), with a fitted slope of 1.00 $\pm$ 0.01 and $\mathrm{R^2}$ = 0.999. The residuals between the set distance and the measured distance indicate high precision. The precision of 0.6 mm is estimated from the standard deviation of the measured distances with ten repeated measurements at each distance.

We further evaluate the optimized precision of our system. By measuring the relative distances between two targets at different positions, we can effectively eliminate the fiber-length fluctuations induced by temperature. Figure \ref{fig:Fig3}(c) displays the results of continuous distance measurements over 1 hour at a distance of 8.045 m, with the acquisition time of 1 s and the precision of 0.78 $\pm$ 0.02 mm. According to Eq. (\ref{eq8}) and Fig. \ref{fig:Fig3}(d), it is indicated that the precision of the demonstrated quantum AMCW LiDAR could be improved with the increase in acquisition time and modulation frequency. Using our system, we could enable the precision to reach 29 $\pm$ 4 $\mathrm{\mu m}$ with the number of correlated photon pairs growing, when the acquisition time is up to 500 s. Meanwhile, the higher modulation frequency will be limited by the total system time jitter mentioned earlier, and higher frequencies will result in lower visibility of the modulated idler photons, thereby reducing precision; see details in the supplementary material.

\subsection*{Precision under noise}
\begin{figure*}[h!]
\centering
\includegraphics[width=12 cm]{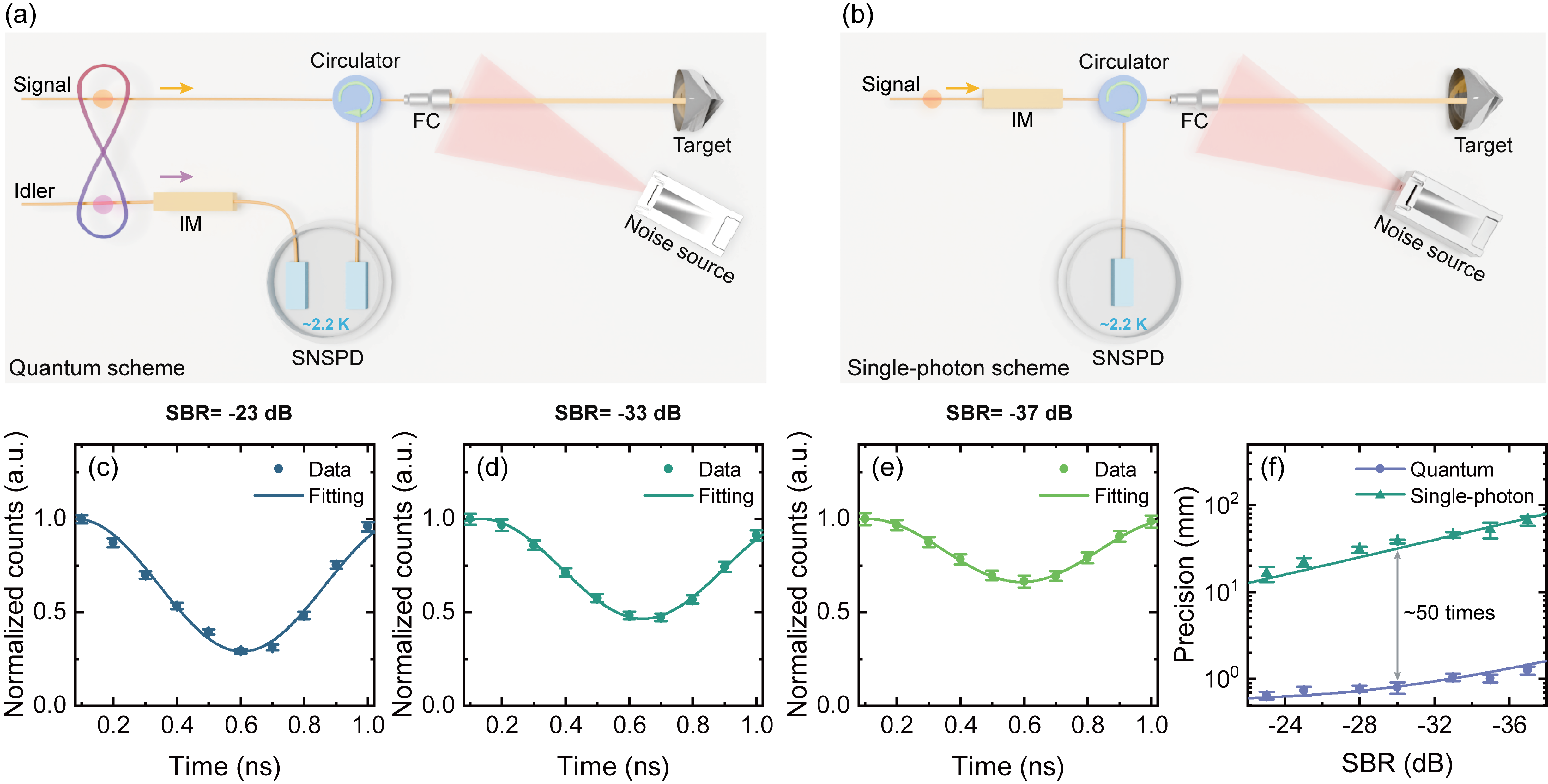}
\caption{Precision of quantum and single-photon schemes under noise. (a) and (b) are the schematics of quantum and single-photon AMCW LiDAR, respectively. (c)-(e) Correlation curves with the following SBR: -23 dB, -33 dB, and -37 dB. To avoid the saturation of SPD, the number of correlated photon pairs is 1.4 kHz, and the acquisition time is 10 s. (f) Comparison of precision between quantum and single-photon schemes versus the SBR. The error bar is calculated from the standard deviation of 1000-time Monte Carlo simulations. The solid line is the theoretical curve of Eq. (\ref{eq8}).}
\label{fig:Fig4}
\end{figure*}

Quantum AMCW LiDAR can offer enhanced precision over single-photon AMCW LiDAR in the presence of strong background noise. To evaluate the noise resilience, we introduce a strong background illumination to the system, which is simulated by using an Erbium-doped fiber light source (EDFLS) with a spectral range of 60 nm at a center wavelength of 1.5 $\mathrm{\mu m}$. This broadband light is coupled into free space via a fiber collimator and mixed with signal photons at a small incident angle. The noise photon flux is quantified by the SNSPD when the signal photons are blocked. For comparison, the single-photon AMCW LiDAR without quantum correlation is implemented at the same illumination level as the quantum scheme, as shown in Fig. \ref{fig:Fig4}(a) and (b). Note that the single-photon AMCW LiDAR uses the modulated single-sided counts of the signal channel to estimate distance.

The difference of precision between quantum ACMW LiDAR and single-photon AMCW LiDAR is performed with various levels of background noise and the same probe photon counts. In the quantum scheme, a narrow time window with 1 ns width is used to select only temporally correlated signal-idler photon pairs, effectively rejecting uncorrelated noise photons outside this window. Figure \ref{fig:Fig4}(c)-(e) shows the correlation curves between the signal and idler photons at SBRs of -23 dB, -33 dB, and -37 dB, respectively. To avoid the saturation of SNSPD, the number of correlated photon pairs is maintained at 1.4 kHz, with an acquisition time of 10 s per data point. As the SBR decreases (i.e., background noise increases), the visibility in the quantum AMCW LiDAR system drops from 0.43 $\pm$ 0.01 to 0.19 $\pm$ 0.01 with the SBR decreasing from -23 dB to -37 dB, and the corresponding precision degrades from 0.64 $\pm$ 0.06 mm to 1.2 $\pm$ 0.1 mm. 

Figure \ref{fig:Fig4}(f) compares the precision of quantum and single-photon methods under different SBR. In the single-photon AMCW LiDAR scheme, the total detected photon counting rate, the acquisition time and the phase estimation method are identical to those of the quantum scheme, and the precision degrades from 29 $\pm$ 3 mm to 66 $\pm$ 8 mm with the same SBR decreasing range, which is degraded more than a factor of 50 compared to the quantum scheme. Remarkably, even if the background noise photon flux exceeds the number of correlated photon pairs by a factor of 5000, the quantum AMCW LiDAR system retains millimeter-level precision at one second, demonstrating its exceptional resilience to the strong background noise.

\section*{Discussion and conclusion}
\begin{figure}[h!]
\centering
\includegraphics[width=9 cm]{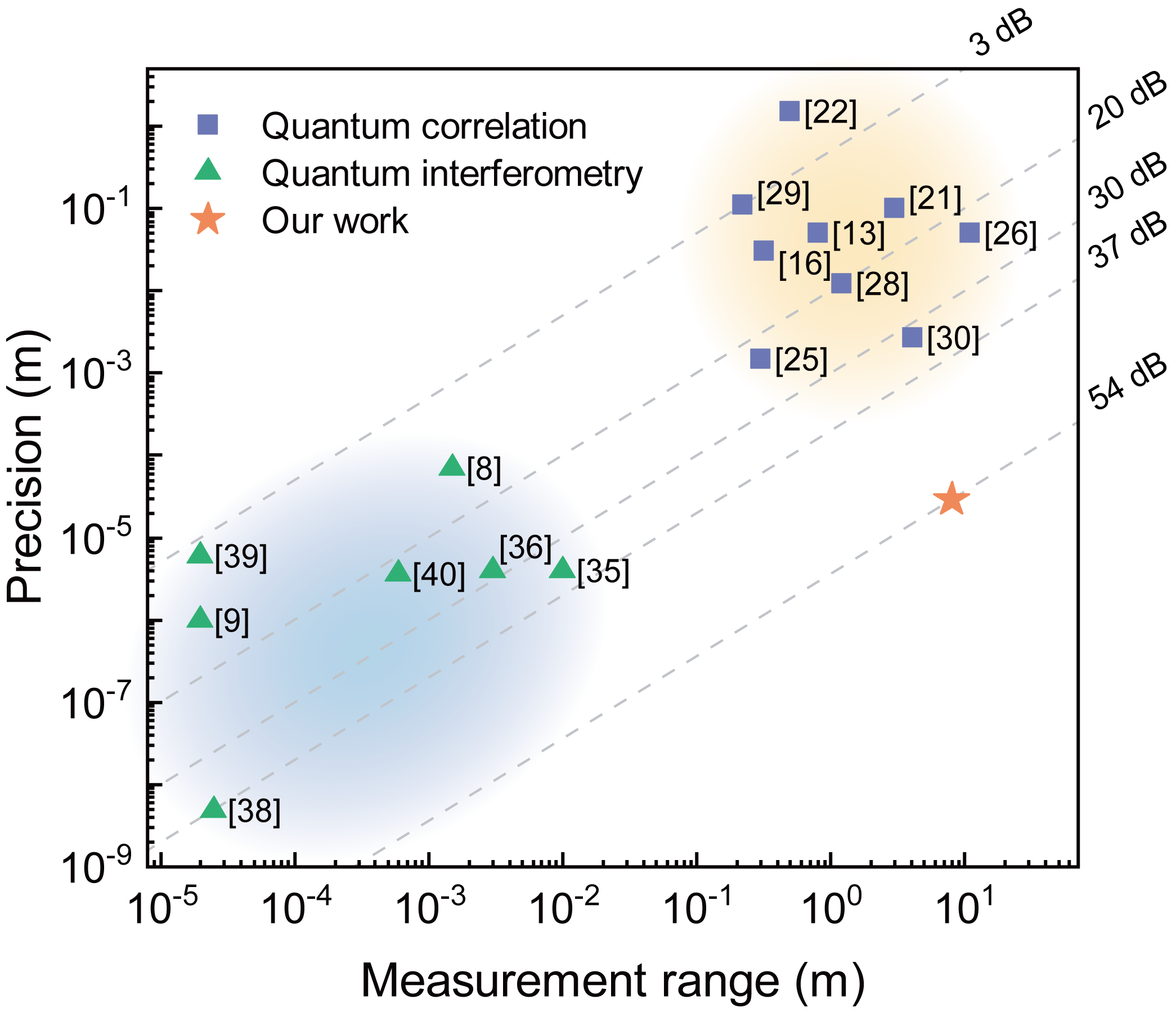}
\caption{Precision versus measurement range for different quantum LiDAR systems. These systems can be classified into two groups: one characterized by high precision (blue shadow) and the other by large measurement range (orange shadow).}
\label{fig:Fig5}
\end{figure}

We demonstrate a quantum LiDAR scheme based on non-local modulation of the correlated photon pairs. Benefiting from non-local modulation and quantum correlation, our system with a total time jitter of 300 ps simultaneously achieves micrometer-level precision, a large measurement range, and high noise tolerance. As shown in Fig. \ref{fig:Fig5}, compared to the previous quantum LiDAR scheme, the dynamic range of quantum AMCW LiDAR, defined as 10log (measurement range/precision), reaches 54 dB, showing its advantage over the other quantum LiDAR systems. Moreover, the measurement range of our quantum AMCW LiDAR could be enlarged effectively as the transmission attenuation decreases in the signal channel, and we achieve millimeter-level precision in a single-mode fiber with a length of 10 km, see the detailed analysis and discussion in the supplementary material.

The performance of quantum AMCW LiDAR surpasses that of conventional AMCW LiDAR without the quantum correlation. Quantum AMCW LiDAR utilizing the inherent quantum correlations between the photon pairs breaks the trade-off between the precision and the measurement range; the measurement range of quantum AMCW LiDAR extends beyond the unambiguous range limited by the modulation frequency. Moreover, the precision of quantum AMCW LiDAR is superior to that of conventional AMCW LiDAR under strong background noise, because uncorrelated noise photons can be eliminated via temporal filtering. Furthermore, quantum AMCW LiDAR offers superior covertness compared to conventional AMCW LiDAR for the following reasons. Firstly, the illumination power of our quantum AMCW LiDAR is 1.8 fW, which is below the power of environmental near-infrared radiation. Secondly, the random temporal distribution of the correlated photons is similar to that of the background noise photons \cite{RN10}. Thirdly, due to the nonlocal correlations inherent in the correlated photon pairs, the object will not be directly illuminated by the modulated photons. All these advantages enable the probe signal photons to be indistinguishable from the environmental noise at the target, making detection or jamming extremely difficult in the covert ranging applications. 

In conclusion, our presented quantum AMCW LiDAR method with a high-brightness and low-noise quantum light source can further support rapid, high-precision quantum ranging even under strong environmental background noise \cite{Yan2026-SA, RN58, RN60, RN59}. If we adjust the wavelength of the quantum light source, our quantum AMCW LiDAR can expand its applications, such as the mid-infrared wavelength can be used for free-space ranging with lower losses; photons in the blue-green wavelength range enable low-loss underwater ranging. This work will promote and accelerate the practical application of quantum LiDAR.

\section*{Materials and Methods}
\subsection*{PPLN module design}
We use a fiber-integrating PPLN waveguide with a length of 50 mm to generate the correlated photon pairs through cascaded nonlinear processes of SHG and SPDC. A few spontaneous Raman scattering noise photons are generated by connecting the PPLN waveguide and noise-rejecting filters with a fiber of 20 cm long. This results in the fact that the correlated photon pairs are obtained with a large number of correlated photon pairs and low noise. More details about the parameters of our PPLN module are listed in Table S1 of supplementary material.

\subsection*{Spectral and spatial filtering methods}
Spectral and spatial filtering techniques are employed to suppress ambient light, which can effectively avoid the saturation of SNSPD. Since natural sunlight contains relatively little power around 1.5 $\mathrm{\mu m}$, a narrow-band fiber filter can be used to reject the majority of out-of-band background noise photons. The echo signal photons are detected by an SNSPD through a narrow-band fiber filter centered at 1531.90 nm, with a spectral full-width at half-maximum (FWHM) of 0.8 nm and a sideband extinction ratio of 120 dB. Furthermore, spatial filtering is implemented by coupling the return signal into a single-mode fiber (SMF), which accepts only photons within a narrow angular cone. The SMF has a field of view (FOV) of 10 $\mathrm{\mu rad}$, effectively rejecting off-axis solar background and stray light. With these combined filtering strategies, the background noise photon count rate can be reduced to 1.6 kHz under indoor illumination (approximately 500 lx).

%%%%%%%%%%%%%%%% REFERENCES %%%%%%%%%%%%%%%

\clearpage
\bibliography{sn-bibliography}
% \bibliographystyle{sciencemag}

%%%%%%%%%%%%%%%% ACKNOWLEDGEMENTS %%%%%%%%%%%%%%%
\clearpage
\section*{Acknowledgments}
\paragraph*{Funding:}
This work was supported by Quantum Science and Technology-National Science and Technology Major Project (Nos. 2024ZD0300800, 2021ZD0300701), Sichuan Science and Technology Program (Nos. 2024YFHZ0370, 2024YFHZ0369, 2024YFHZ0368), National Natural Science Foundation of China (Nos. 62475039, 62405046, 62375043, 92365106), Tianfu Jiangxi Laboratory (No. TFJX-ZD-2025-004)

\paragraph*{Competing interests:}
Authors declare that they have no competing interests.

\paragraph*{Data and materials availability:}
Data underlying the results presented in this paper are not publicly available at this time but may be obtained from the authors upon reasonable request.
%%%%%%%%%%%%%%%% END OF MAIN TEXT %%%%%%%%%%%%%%%
\newpage
\begin{appendices}
\section*{Note S1: Derivation of quantum AMCW LiDAR}
The two-photon state generated in the SPDC process could be expressed as \cite{RN55}:
\begin{equation}
    |\Psi \rangle =\int_{-\infty }^{\infty }{d{{t}_{s}}d}{{t}_{i}}\delta ({{t}_{s}}-{{t}_{i}})\hat{a}_{s}^{\dagger }\left( {{t}_{s}} \right)\hat{a}_{i}^{\dagger }\left( {{t}_{i}} \right)|vac\rangle
\end{equation}

where, the indices $s$ and $i$ indicate the signal photons (with time $t_s$) and idler photons (with time $t_i$), $\hat{a}_{s}^{\dagger }$ and $\hat{a}_{i}^{\dagger }$ are the creation operator of signal and idler photons at time ${{t}_{s}}$ and ${{t}_{i}}$. In the frequency domain, the two-photon state can be rewritten as:
\begin{equation}
    |\Psi \rangle =\int_{-\infty }^{\infty }{dvf(v)}\hat{a}_{s}^{\dagger }\left( \omega _{s}^{0}+v \right)\hat{a}_{i}^{\dagger }\left( \omega _{i}^{0}-v \right)|vac\rangle
\end{equation}

where $\omega _{s}^{0}$ and $\omega _{i}^{0}$ are the central angular frequencies of the signal and idler photons, $v$ is the angular frequency shift, and $f(v)$ is the spectral distribution of correlated photon-pairs. In the measurement arm, the signal photons are transmitted to the target and reflected. The positive-field operator of the signal photons can be expressed as
\begin{equation}
    E_{s}^{\left( + \right)}\left( {{t}_{s}} \right)=\int_{-\infty }^{\infty }{d{{\omega }_{s}}\sqrt{{{\alpha }_{s}}}{{{\hat{a}}}_{s}}\left( {{\omega }_{s}} \right){{e}^{-i{{\omega }_{s}}({{t}_{s}}+\tau )}}}
\end{equation}
where, ${{\omega }_{s}}$ is the angular frequency of the signal photons, ${{\omega }_{s}}$ is the transmission of signal photons at measuring arm, ${{\hat{a}}_{s}}\left( {{\omega }_{s}} \right)$ is the annihilation operator of the signal photons at the frequency ${{\omega }_{s}}$, and $\tau $ is the round-trip delay of signal photons and equals to $2d/c$ ($c$ is the speed of light in vacuum). In the herald arm, the temporal distribution of idler photons is modulated by an intensity modulator, and the idler photons are detected by the single-photon detector. Then, the positive-field operator of the idler photons at time ${{t}_{i}}$ can be expressed as
\begin{equation}
    E_{i}^{\left( + \right)}\left( {{t}_{i}} \right)=\int_{-\infty }^{\infty }{d{{\omega }_{i}}\sqrt{\frac{{{\alpha }_{i}}}{2}\left[ 1+V\cos \,\left( 2\pi {{f}_{m}}{{t}_{i}}+{{\varphi }_{0}} \right) \right]}{{{\hat{a}}}_{i}}\left( {{\omega }_{i}} \right){{e}^{-i{{\omega }_{i}}{{t}_{i}}}}}
\end{equation}
where, ${{\omega }_{i}}$ is the frequency of the idler photons, ${{\alpha }_{i}}$ is the transmission of idler photons at herald arm, ${{\hat{a}}_{i}}\left( {{\omega }_{i}} \right)$ is the annihilation operator of the idler photons at the frequency ${{\omega }_{i}}$, ${{f}_{m}}$ is the modulation frequency, $V$ is the visibility of modulation signal, and ${{\varphi }_{0}}$ is the initial phase. 

In the biphoton state, the probability for jointly observing a pair of photons can be measured by the second-order correlation function, which can be calculated from \cite{RN54}:
\begin{equation}
    {{G}^{\left( 2 \right)}}\left( {{t}_{1}},{{t}_{2}} \right)=\left\langle  \Psi  \right|E_{s}^{\left( - \right)}\left( {{t}_{s}} \right)E_{i}^{\left( - \right)}\left( {{t}_{i}} \right)E_{s}^{\left( + \right)}\left( {{t}_{s}} \right)E_{i}^{\left( + \right)}\left( {{t}_{i}} \right)\left| \Psi  \right\rangle
\end{equation}
The wavefunction of a photon pair emitted from the nonlinear crystal is 
\begin{equation}
    \Psi \left( {{t}_{s}},{{t}_{i}} \right)=\left\langle  0 \right|E_{s}^{\left( + \right)}\left( {{t}_{s}} \right)E_{i}^{\left( + \right)}\left( {{t}_{i}} \right)\left| \Psi  \right\rangle
\end{equation}
Then, the second-order correlation function is written as
\begin{equation}
    {{G}^{\left( 2 \right)}}\left( {{t}_{1}},{{t}_{2}} \right)={{\left| \left\langle  0 \right|E_{s}^{\left( + \right)}\left( {{t}_{1}} \right)E_{i}^{\left( + \right)}\left( {{t}_{2}} \right)\left| \Psi  \right\rangle \text{ } \right|}^{2}}
\end{equation}
Substituting Eqs. (1-4) into Eq. (6), the wavefunction can be written as
\begin{equation}
    \Psi \left( {{t}_{s}},{{t}_{i}} \right)=\sqrt{\frac{1}{2}{{\alpha }_{s}}{{\alpha }_{i}}\left[ 1+V\cos \,\left( 2\pi {{f}_{m}}{{t}_{i}}+{{\varphi }_{0}} \right) \right]}{{e}^{-i\left( \omega _{s}^{0}{{t}_{s}}+\omega _{i}^{0}{{t}_{i}} \right)}}{{\mathcal{F}}_{{{t}_{s}}-{{t}_{i}}}}\left\{ f\left( v \right) \right\}
\end{equation}
where, ${{\mathcal{F}}_{{{t}_{s}}-{{t}_{i}}}}\left\{ f\left( v \right) \right\}$ is the Fourier transform of $f\left( v \right)$. Generally, the $f\left( v \right)$ is wide spectral distribution, and the ${{\mathcal{F}}_{{{t}_{s}}-{{t}_{i}}}}\left\{ f\left( v \right) \right\}$ can be written as a delta-like function. Then, the second-order correlation function is expressed as
\begin{equation}
    {{G}^{\left( 2 \right)}}\left( {{t}_{s}},{{t}_{i}},\tau  \right)=\frac{1}{2}{{\alpha }_{s}}{{\alpha }_{i}}\left[ 1+V\cos \,\left( 2\pi {{f}_{m}}{{t}_{i}}+{{\varphi }_{0}} \right) \right]\delta \left( {{t}_{i}}-{{t}_{s}}-\tau  \right)
\end{equation}
Furthermore, the time of idler photon ${{t}_{i}}$ can be replaced by that of signal photons ${{t}_{s}}$. Thus, the second-order correlation function can be rewritten as:
\begin{equation}
    {{G}^{\left( 2 \right)}}\left( {{t}_{s}},\tau  \right)=\frac{1}{2}{{\alpha }_{s}}{{\alpha }_{i}}\left[ 1+V\cos \left( 2\pi {{f}_{m}}{{t}_{s}}+\Delta \varphi +{{\varphi }_{0}} \right)\, \right]
\end{equation}
where, the $\Delta \varphi =4\pi {{f}_{m}}d/c$ is the phase shifting caused by the target displacement. Furthermore, the number of correlated photon pairs can be measured by using the coincidence measurement. The number of correlated photon pairs, named coincidence counts ${{N}_{cc}}$, is the counts of Eq. (S9) that fall into a certain time window around a central time ${{t}_{i}}-{{t}_{s}}=\tau$. After the sampling of the time-digital converter (TDC), the temporal discrete distribution of correlated counts accumulated over a modulation period can be written as 
\begin{equation}
    {{N}_{k}}=\frac{{{N}_{cc}}}{{{f}_{m}}\Delta {{t}_{bin}}}\left[ 1+V\cos (2\pi {{f}_{m}}{{t}_{k}}+\Delta \varphi +{{\varphi }_{0}}) \right]
\end{equation}
where, $\Delta {{t}_{bin}}$ is the bin width of TDC, $k$ (= 1, 2, …, $K$) is the sequence of discrete samples.

\newpage
\section*{Note S2: Phase estimation and Cramer-Rao Bound of quantum AMCW LiDAR}
The temporal distribution of correlation photons sampled by the TDC can be simplified to
\begin{equation}
    {{N}_{k}}=a\cos (2\pi {{f}_{m}}{{t}_{k}}+\varphi )+b
\end{equation}
where $a=V{{N}_{cc}}/K$, $b={{N}_{cc}}/K$, and $K={{f}_{m}}\Delta {{t}_{bin}}$ is the number of sampling points in a modulation period. The temporal distribution of correlated counts ${{N}_{k}}$ follows the Poisson distribution, and the probability distribution function of ${{N}_{k}}$ can be given by:
\begin{equation}
    P({{N}_{k}};\varphi )=\frac{{{{\bar{N}}}_{k}}^{{{N}_{k}}}{{e}^{-{{{\bar{N}}}_{k}}}}}{{{N}_{k}}!}
\end{equation}
where, ${{\bar{N}}_{k}}$ is the average value of ${{N}_{k}}$ at each sampling point. Then, the phase can be estimated by the maximal likelihood estimation, and the log-likelihood function is given by: 
\begin{equation}
    \mathcal{L}(\varphi )\propto -\frac{1}{2b}\sum\limits_{k=1}^{K}{{{\left[ {{N}_{k}}-(a\cos (2\pi {{f}_{m}}{{t}_{k}}+\varphi )+b) \right]}^{2}}}
\end{equation}
The maximized log-likelihood function can be solved by minimizing the squared error: 
\begin{equation}
    \min J(\varphi )=\sum\limits_{k=1}^{K}{{{\left[ {{N}_{k}}-b-a\cos (2\pi {{f}_{m}}{{t}_{k}}+\varphi ) \right]}^{2}}}
\end{equation}
The estimated phase in a period can be solved when$J(\varphi )=0$:
\begin{equation}
    \hat{\varphi }=\text{atan}\left( -\frac{\sum\limits_{k=0}^{K-1}{(}{{N}_{k}}-b)\sin (2\pi {{f}_{m}}{{t}_{k}})}{\sum\limits_{k=0}^{K-1}{(}{{N}_{k}}-b)\cos (2\pi {{f}_{m}}{{t}_{k}})} \right)
\end{equation}

The ultimate limit on the precision of estimated phase is determined by the Cramér–Rao bound, which states that the variance of any unbiased estimator must be bounded by \cite{RN32}
\begin{equation}
    {{\sigma }^{2}}\hat{\varphi }\ge \frac{1}{I\left( \varphi  \right)}
\end{equation}
The Fisher information measures the amount of information about $\hat{\varphi }$. The Fisher information of ${{N}_{k}}$ is calculated by the joint probability distribution of photon counts:
\begin{equation}
    {{I}_{k}}=p\left( {{N}_{k}};\varphi  \right){{\left( \frac{\partial }{\partial \varphi }\log p\left( {{N}_{k}};\varphi  \right) \right)}^{2}}
\end{equation}

Then, the total Fisher information can be given by
\begin{equation}
    I\left( \varphi  \right)=\sum\limits_{k=1}^{K}{{{I}_{k}}}={{a}^{2}}\sum\limits_{k=1}^{K}{\frac{{{\sin }^{2}}\,\left( 2\pi {{f}_{m}}{{t}_{k}}+\varphi  \right)}{a\cos \,\left( 2\pi {{f}_{m}}{{t}_{k}}+\varphi  \right)+b}}
\end{equation}
In a period of modulated signal, $\sum\limits_{k=1}^{K}{{{\sin }^{2}}\,\left( 2\pi {{f}_{m}}{{t}_{k}}+\phi  \right)}\approx \frac{K}{2}$. Thus, $I\left( \varphi  \right)$ can be simplified as
\begin{equation}
    I\left( \varphi  \right)=\frac{{{a}^{2}}K}{2b}=\frac{{{V}^{2}}{{N}_{cc}}}{2}
\end{equation}
Thus, the Cramer-Rao bound of the estimated phase of correlated photon pairs can be expressed as 
\begin{equation}
    CRB=\frac{\sqrt{2}}{V\sqrt{{{N}_{cc}}}}
\end{equation}
Furthermore, the minimum uncertainty of the estimated distance is deduced by:
\begin{equation}
    \sigma (\hat{d})=\frac{c}{4\pi {{f}_{m}}}CRB=\frac{c\sqrt{2}}{4\pi {{f}_{m}}V\sqrt{{{N}_{cc}}}}
\end{equation}

\begin{figure*}[h!]
\centering
\includegraphics[width=10 cm]{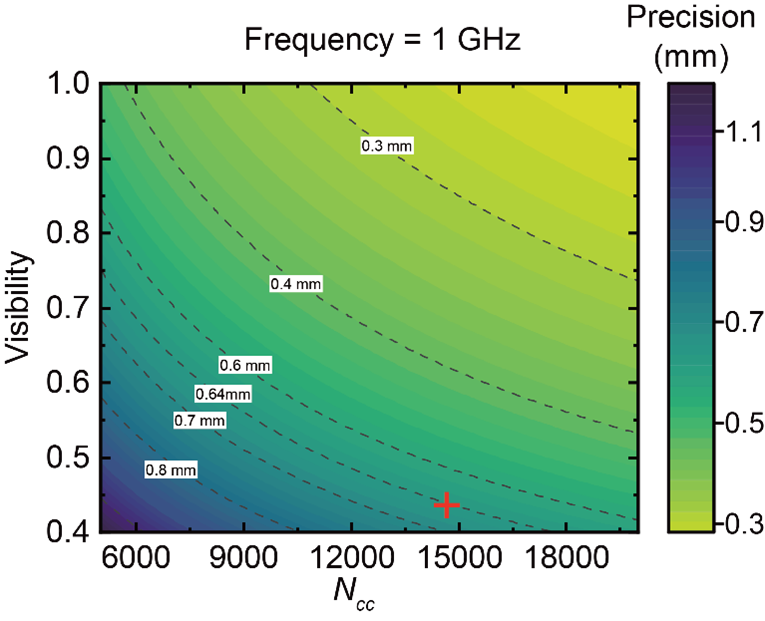}
\caption{Theoretical precision of quantum AMCW LiDAR. ‘+’ shows the parameters in our experiment with the visibility of 0.43 and the $N_{cc}$ of 14 kHz.}
\label{fig:FigS1}
\end{figure*}

\newpage
\section*{Note S3: Theoretical precision of quantum AMCW LiDAR under noise}
Under the noise, the accidental coincidence counts between the noise and idler photons in the time window are calculated as
\begin{equation}
    {{N}_{ac}}={{N}_{b}}{{N}_{i}}\Delta {{\tau }_{w}}
\end{equation}
Through the temporal correlation, the correlation curves contain accidental coincidence counts, and the temporal distribution of photon counts can be expressed as:
\begin{equation}
    {{N}_{k}}=\frac{V{{N}_{cc}}}{{{f}_{m}}\Delta {{t}_{bin}}}\cos (2\pi {{f}_{m}}{{t}_{k}}+\varphi )+\frac{{{N}_{cc}}+{{N}_{ac}}}{{{f}_{m}}\Delta {{t}_{bin}}}
\end{equation}
Then, the visibility can be rewritten as
\begin{equation}
    {V}'=V\sqrt{\frac{{{N}_{cc}}}{{{N}_{cc}}+{{N}_{b}}{{N}_{i}}\Delta {{\tau }_{w}}}}
\end{equation}
According to Section S2 above, the minimum uncertainty of the estimated distance under noise can be deduced by
\begin{equation}
    \sigma (\hat{d})=\frac{c\sqrt{2}}{4\pi {{f}_{m}}V\sqrt{{{N}_{cc}}}}\sqrt{\frac{{{N}_{cc}}+{{N}_{b}}{{N}_{i}}\Delta {{\tau }_{w}}}{{{N}_{cc}}}}
\end{equation}
The signal-to-background noise ratio (SBR) is the ratio between the number of correlated photon pairs and that of noise photons. Thus, the uncertainty can be rewritten as
\begin{equation}
    \sigma (\hat{d})=\frac{\sqrt{2}}{4\pi {{f}_{m}}V\sqrt{{{N}_{cc}}}}\sqrt{1+\frac{{{N}_{i}}\Delta {{\tau }_{w}}}{SBR}}
\end{equation}

\newpage
\section*{Note S4: Generation of correlated photon-pairs in a periodically poled lithium niobate (PPLN) waveguide}
We use a fiber-integrating PPLN waveguide with a length of 50 mm to generate the correlated photon pairs through cascaded nonlinear processes of second harmonic generation (SHG) and spontaneous parameter down-conversion (SPDC) \cite{RN45}, shown in Fig. S2. A few spontaneous Raman scattering noise photons are generated by connecting the PPLN waveguide and noise-rejecting filters with a fiber of 20 cm long. This results in the fact that the correlated photon pairs can be obtained with a high generation rate and low noise. The parameters of the PPLN module are shown in Table S1.
\begin{figure*}[h!]
\centering
\includegraphics[width=7 cm]{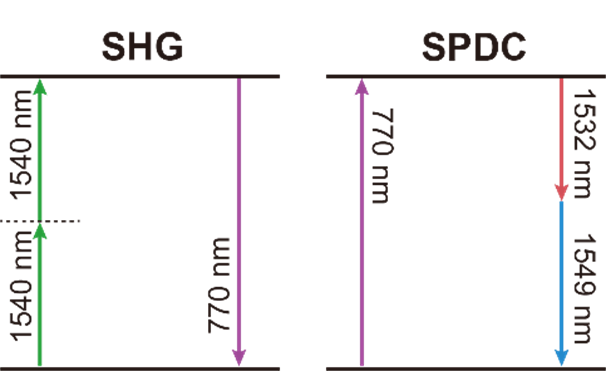}
\caption{Schematic illustration of the SHG and SPDC processes.}
\label{fig:FigS2}
\end{figure*}

\begin{table}[htbp]
\centering
\tabcolsep=1.2 cm
\renewcommand\arraystretch{1}
\caption{Parameters of PPLN module.}
\begin{tabular}{cc}
\hline
Type of waveguide & RPE waveguide\\
Length of waveguide & 50 mm\\
QPM period & 19 $\mathrm{\mu m}$\\
SHG normalized conversion efficiency & 500\%@1540.56 nm\\
Length of pigtail & 20 cm\\
Input coupling efficiency of PPLN waveguide & 73.7\%\\
Output coupling efficiency of PPLN waveguide & 85.7\%\\
\hline
\end{tabular}
\end{table}
Figure S3 is the experimental setup to generate the correlated photon pairs. A continuous wave laser (PPCL300, PURE Photonics) with a linewidth of 100 kHz at 1540.32 nm pumps a PPLN waveguide module with a temperature controller to generate the correlated photon pairs. To overcome the losses of the intensity modulator (IM) and the imaging path, an Erbium-doped fiber amplifier (EDFA) module is used to boost the optical power of the pump laser. A fiber beam splitter with a ratio of 99:1 and a power meter (PM) are used to adjust and monitor the laser power. A polarization controller (PC) is used to ensure polarization alignment to maximize the efficiency of phase matching in the PPLN waveguide. 

\begin{figure*}[h!]
\centering
\includegraphics[width=12 cm]{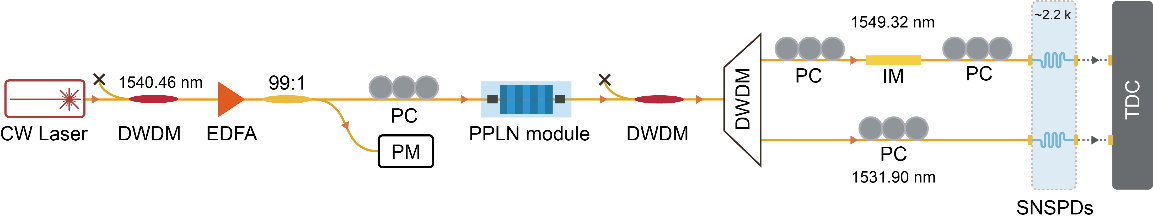}
\caption{Experimental setup for the generation of correlated photon pairs.}
\label{fig:FigS3}
\end{figure*}

\begin{table}[htbp]
\centering
\tabcolsep=0.6 cm
\renewcommand\arraystretch{1}
\caption{Parameters of SNSPD.}
\begin{tabular}{cccc}
\hline
Channel & Efficiency (\%) & Dark count rate (Hz) & Time jitter (ps)\\
\hline
Signal & 74\% & 45 & 234\\
Idler & 80\% & 38 & 238\\
\hline
\end{tabular}
\end{table}

A dense wavelength division multiplexer (DWDM) with a central reflection wavelength at 1540.56 nm filters the residual pump laser. The entangled photon pairs are spectrally filtered as signal photons (1531.90 nm) and idler photons (1549.32 nm) using two cascaded DWDMs with a bandwidth of 100 GHz. The signal photons are directly transmitted to a superconducting nanowire single-photon detector (SNSPD, P-CS- 6, PHOTEC). The idler photons are modulated by the intensity modulator and then are detected by the SNSPD. A polarization controller (PC) is used to ensure polarization alignment to maximize SNSPD efficiency. The output electrical signals of SNSPDs are recorded by the time-to-digital converter (TDC, ID Quantique, ID900) to perform the correlation measurement. The parameters of SNSPD are shown in Table S2. The coincidence counts and CAR, defined as the ratio of coincidence counts to accidental coincidence counts, are used to evaluate the distinguishability of coincidence peaks. The coincidence counts between the signal and idler photons are the total counts within a coincidence time duration covering the coincidence peak, and the accidental coincidence counts between the noise and idler photons are the total counts within the same coincidence time duration away from the coincidence peak. In our experiment, the CAR of 20 and the coincidence counts of 20 k can be obtained.
\newpage
\section*{Note S5: Correlation curves with different time window width}
In the temporal correlation analysis, the time window width can be used to filter out the noise photons, and the visibility of correlation curves can be improved. However, the uncorrelated curves between the noise and idler photons influence the demodulated phase and result in the systematic error that causes the measured distance to differ from the set distance. When the time window width equals an integer multiple of the modulation period, the reconstructed curves between the noise and idler photons are accumulated as a flat waveform, and the phase interference of noise photons is eliminated. As shown in Fig. S4, for a modulation frequency of 1 GHz, when the time window width is set as 0.4 ns, the uncorrelation curves between the noise and idler photons can be observed, and the phase shift varies continuously with time delays. In the time window including the correlated photon pairs, the reconstructed curves are the sum of the correlation and uncorrelation curves, and the phase undergoes a discontinuity. When the time window width is set as 1 ns, the uncorrelation curves between the noise and idler photons are flat waveforms. Only the temporal range including the correlated photon pairs, the reconstructed curves with a cosine shape are observed. Furthermore, for an optimal noise resilience, the time window width should be set as the modulation period to filter out the noise photons.
\newpage
\begin{figure*}[h!]
\centering
\includegraphics[width=12 cm]{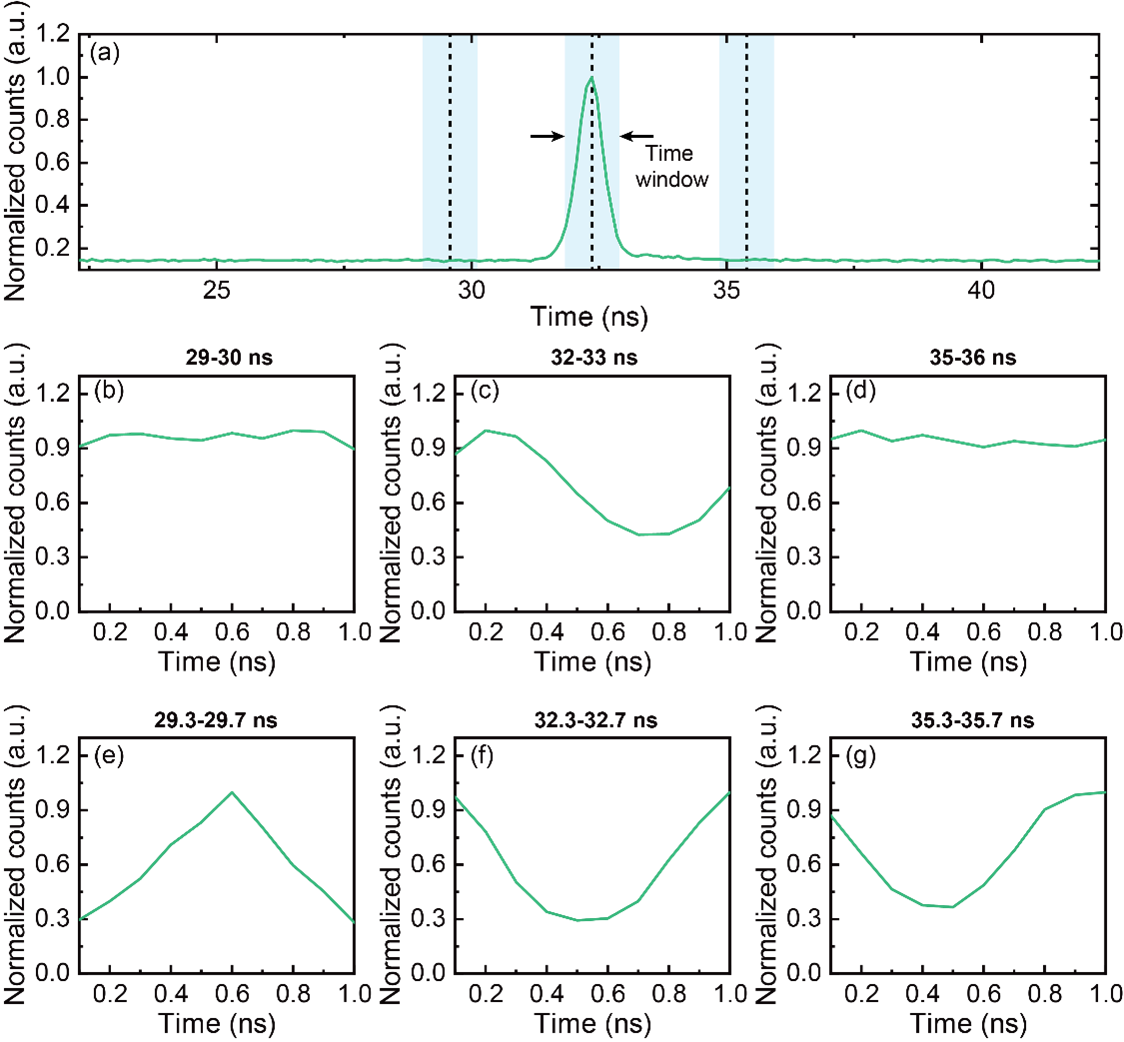}
\caption{Reconstructed curves of correlated photon pairs and noise under different time window widths. (a) The reconstructed curves between idler and signal photons. (b)-(d) are the reconstructed curves over a 1-ns modulation period when the time window width is 1 ns. (e)-(g) are the reconstructed curves over a 1-ns modulation period when the time window width is 0.4 ns.}
\label{fig:FigS4}
\end{figure*}

\newpage
\section*{Note S6: Precision under different modulated frequency}
The idler photons from the correlated photon pairs are sinusoidally modulated in intensity by an electro-optic intensity modulator, as illustrated in Fig. S5. An RF bias voltage is applied to the microwave source via an RF bias tee, enabling the adjustment of the DC offset of the microwave signal. A DC bias voltage is used to control and adjust the modulation operating point of the electro-optic intensity modulator. By optimizing the RF bias voltage, the output power of the microwave source (MS, R\&S, SMA100B), and the DC bias voltage of intensity modulation, the visibility of the correlation curves could be maximized.

\begin{figure*}[h!]
\centering
\includegraphics[width=12 cm]{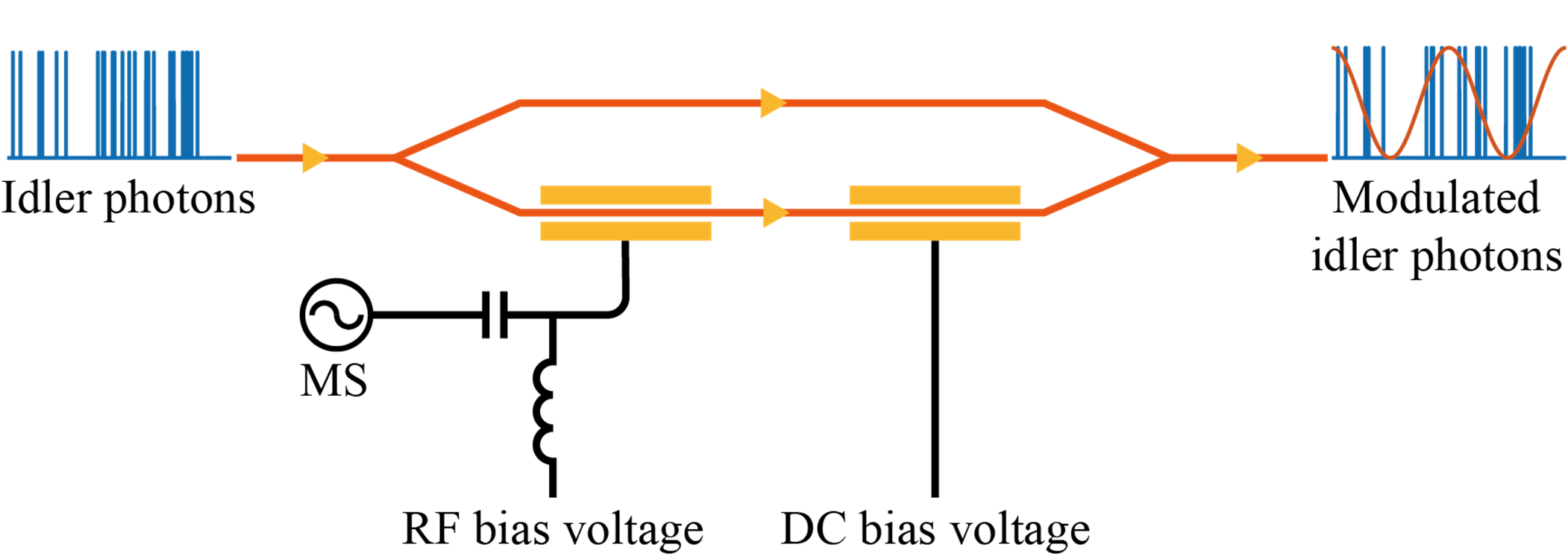}
\caption{Illustrations of intensity modulation for the idler photons.}
\label{fig:FigS5}
\end{figure*}

The reconstructed correlation curve is further degraded by the system’s timing jitter. After long-term accumulation, the measured correlation curve corresponds to the convolution of the ideal modulation waveform with the system’s timing jitter \cite{Wu2019-Optica}. As the modulation frequency increases, the finite bandwidth of the detection system reduces its frequency response, thereby decreasing the visibility of the correlation curves. According to Eq. (S22), the ranging precision depends on the modulation frequency, the fringe visibility, and the coincidence count rate. The temporal window width is set equal to one period of the modulation signal to ensure unambiguous phase extraction. Consequently, the coincidence count rate decreases with the increasing of modulation frequency, as fewer photon pairs fall within the narrower time window.

We measure the modulation curves of correlated photon pairs at different modulation frequencies of 250 MHz, 500 MHz, 1 GHz, and 2 GHz, yielding fringe visibilities of 0.69 $\pm$ 0.01, 0.47 $\pm$ 0.01, 0.43 $\pm$ 0.01, and 0.12 $\pm$ 0.03, respectively, as shown in the Table. S3. The corresponding ranging precisions are summarized, and the best precision of 0.6 mm is achieved at 1 GHz, which represents the optimal trade-off between higher modulation frequency and sufficient photon statistics. Moreover, the 1-ns temporal window at this frequency provides enhanced suppression of uncorrelated background noise.

\begin{table}[htbp]
\centering
\tabcolsep=0.8 cm
\renewcommand\arraystretch{1.2}
\caption{Precision of our quantum AMCW LiDAR under different modulation frequencies.}
\begin{tabular}{ccc}
\hline
Modulation frequency (GHz)	& Visibility & Precision (mm)\\
\hline
0.25 & 0.69 $\pm$ 0.01 & 2.2\\
0.5 & 0.47 $\pm$ 0.01 & 1\\
\textbf{1} & \textbf{0.43 $\pm$ 0.01} & \textbf{0.6}\\
2 & 0.12 $\pm$ 0.03 & 2.4\\
\hline
\end{tabular}
\end{table}

\newpage
\section*{Note S7: Mitigation of fiber-length fluctuations induced by temperature}
For a single target, the absolute distance will be impacted by the fiber-length fluctuations induced by temperature. Figure S6 shows the distance fluctuations of the target and the temperature fluctuations over 10 hours, which exhibit the same temporal variation. To mitigate the fiber-length fluctuations induced by temperature, a reference target is placed, and the relative distance between two targets is measured. Since the returned photons from the two targets' paths through the same transmission fiber have an identical propagation delay in the fiber, as shown in Fig. S6, by subtracting the two measured distances, the result represents the true relative separation between the two targets in free space. This method effectively eliminates the influence of fiber-length variations induced by temperature.

\begin{figure*}[h!]
\centering
\includegraphics[width=12 cm]{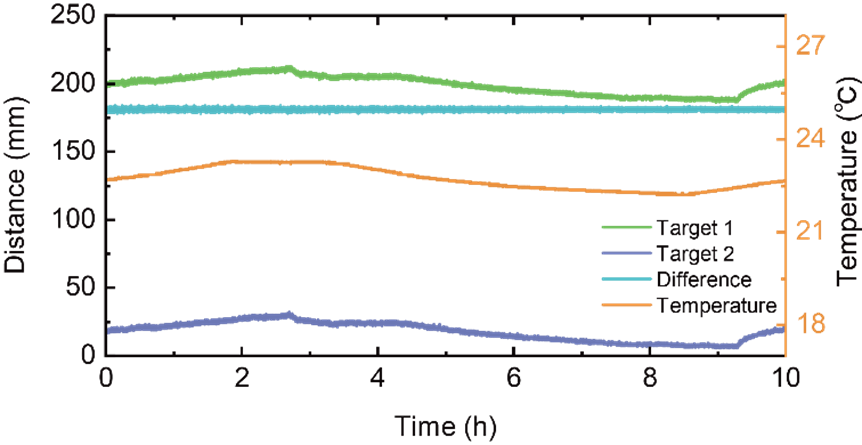}
\caption{Measured distances of two targets and environmental temperature variation over 10 hours.}
\label{fig:FigS6}
\end{figure*}

\newpage
\section*{Note S8: Comparison between quantum and single-photon AMCW LiDAR}

Figure 4(a) and (b) show the experiment setups of quantum and single-photon AMCW LiDAR, respectively. The coincidence counts and single-photon counts are both set as 1.4 kHz, and the acquisition time is set as 10s. In the quantum scheme, the SBR is set to be -23 dB, -33 dB, and -37 dB, respectively, and the visibility in the quantum system drops from 0.43 $\pm$ 0.01 to 0.19 $\pm$ 0.01. The corresponding precision degrades from 0.64 $\pm$ 0.06 mm to 1.2 $\pm$ 0.1 mm. In the single-photon scheme, the noise cannot be filtered out through the temporal correlation. When the SBR is set to be 10 dB, 0 dB, and -10 dB, respectively, and the visibility in the single-photon system drops from 0.32 $\pm$ 0.02 to 0.044 $\pm$ 0.007, as shown in Fig. S7. The corresponding precision degrades from 1.2 $\pm$ 0.6 mm to 3.2 $\pm$ 0.5 mm. Quantum- AMCW LiDAR maintains millimeter-level precision even under strong background noise, whereas conventional single-photon AMCW LiDAR achieves the same precision only when the signal-photon level far exceeds the background noise.

\begin{figure*}[h!]
\centering
\includegraphics[width=12 cm]{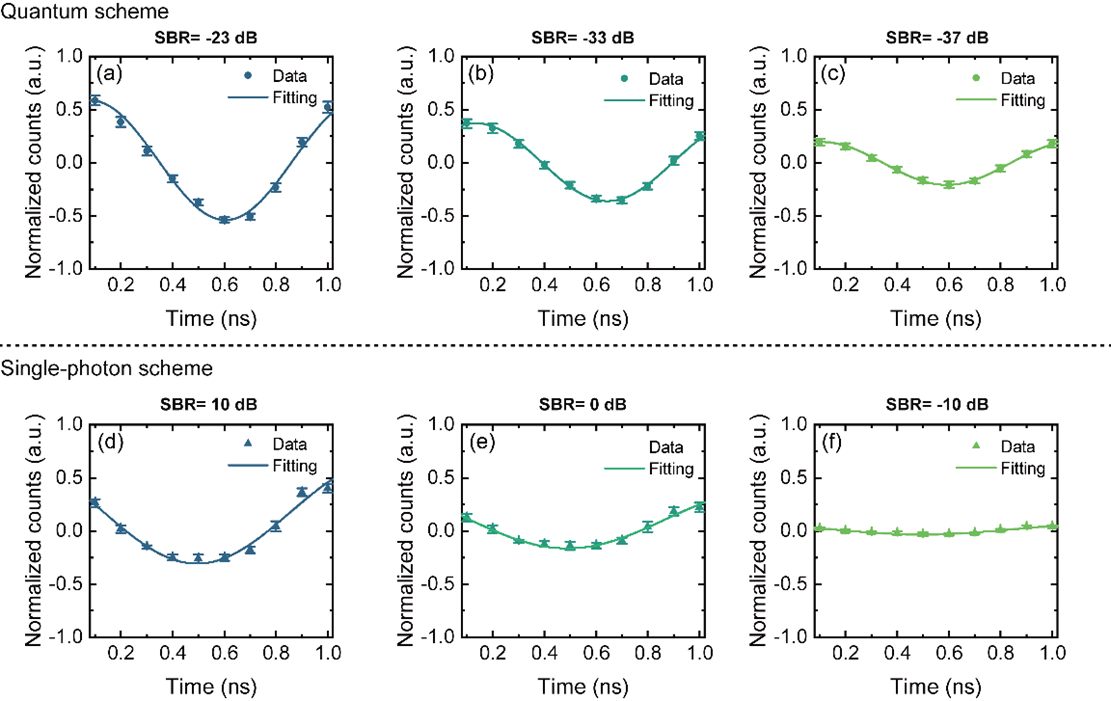}
\caption{Modulation curves of quantum and single-photon AMCW LiDAR. (a)-(c) are the correlation curves of quantum AMCW LiDAR when the SBR is -23 dB, -33 dB, and -37 dB. (d)-(f) are the modulation curves of single-photon AMCW LiDAR when the SBR is 10 dB, 0 dB, and -10 dB.}
\label{fig:FigS7}
\end{figure*}

\newpage
\section*{Note S9: Quantum ranging in the 10-km fiber}

Due to the loss in the free-space channel and the limitation of the experimental site, we demonstrate the long-range quantum AMCW ranging in a 10-km long single-mode fiber (SMF). Figure S8(a) presents the scheme of the experimental setup. The fiber is connected to the output port of the circulator, and the reflected signal photons are detected by the SNSPD. The modulation frequency is set as 250 MHz, generated from an arbitrary function generator (AFG). The end of the under test fiber is connected to a Faraday mirror to improve the end-face reflectivity. With an acquisition time of 1800 s, the correlation curves between the reflected signal and idler photons are shown in Fig. S8(b). The first peak is the reflection of the front end-face of the fiber, and the second peak is the reflection of the Faraday mirror with a length of 1 m. Then, the correlation curves for each peak can be reconstructed, respectively, as shown in Fig. S8(c) and (d). The distance from the fiber end face to the Faraday mirror is determined to be 10091.929 $\pm$ 0.005 m by analyzing the separation between the two coincidence peaks and the corresponding phase of the correlation curves.

\begin{figure*}[h!]
\centering
\includegraphics[width=12 cm]{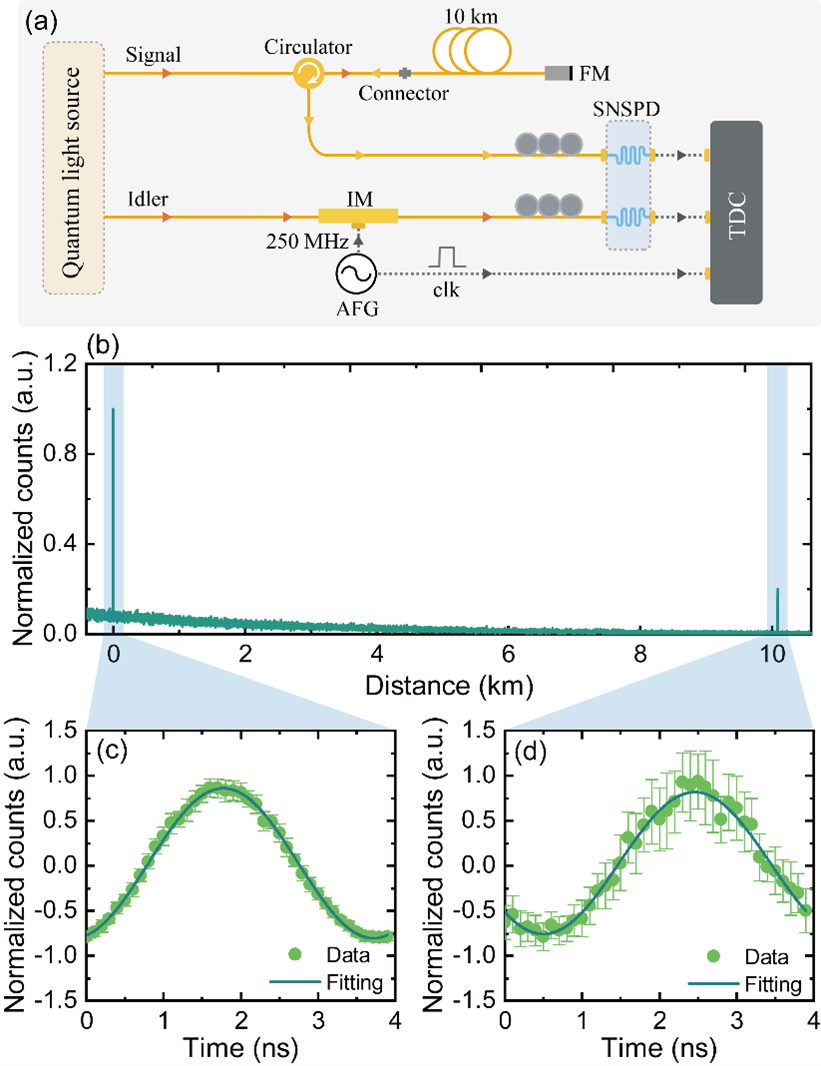}
\caption{Results of quantum AMCW ranging in a 10-km-long SMF. (a) Experimental setup. FM, Faraday mirror; IM, intensity modulator; SNSPD, superconducting nanowire single-photon detectors; AFG, arbitrary function generator; TDC, time-to-digital converter. (b) Coincidence histogram between signal and idler photons with the bin width of 5 ns. The first peak is the reflection of the front end-face of the fiber, and the second peak is the reflection of the Faraday mirror. (c) and (d) are the reconstructed correlation curves of the first and the second peak over one modulated period, respectively. Error bars are calculated by the standard deviation of 1000-time Monte Carlo simulations.}
\label{fig:FigS8}
\end{figure*}

\end{appendices}
\end{document}